\documentclass{article}

\usepackage[final]{nips_2018}

\usepackage[utf8]{inputenc} 
\usepackage[T1]{fontenc}    
\usepackage{hyperref}       
\usepackage{url}            
\usepackage{booktabs}       
\usepackage{amsfonts}       
\usepackage{nicefrac}       
\usepackage{microtype}      
\usepackage{graphicx}       
\usepackage{subcaption}

\title{An AI-based, Multi-stage detection system of banking botnets}

\author{
  Li Ling \\
  JPMorgan Chase\\
  575 Washington Blvd\\
  Jersey City, NJ, 07310, United States \\
  \texttt{li.ling@jpmorgan.com} \\
  \And
  Zhiqiang Gao \\
  JPMorgan Chase\\
  575 Washington Blvd\\
  Jersey City, NJ, 07310, United States \\
  \texttt{zhiqiang.gao@jpmchase.com} \\
  \And
  Michael A Silas  \\
  JPMorgan Chase\\
  575 Washington Blvd\\
  Jersey City, NJ, 07310, United States \\
  \texttt{michael.a.silas@jpmchase.com} \\ 
  \And
  Ian Lee  \\
  JPMorgan Chase\\
  1 Changi Business Park Central\\
  Singapore, 486036, Singapore \\
  \texttt{hock.hi.lee@jpmchase.com} \\ 
  \And
  Erwan A Le Doeuff \\
  JPMorgan Chase\\
  575 Washington Blvd\\
  Jersey City, NJ, 07310, United States \\
  \texttt{erwan.ledoeuff@jpmorgan.com} \\
}

\begin{document}

\maketitle

\begin{abstract}
  Banking Trojans, botnets are primary drivers of financially-motivated cybercrime. In this paper, we first analyzed how an APT-based banking botnet works step by step through the whole lifecycle. Specifically, we present a multi-stage system that detects malicious banking botnet activities which potentially target the organizations. The system leverages Cyber Data Lakes as well as multiple artificial intelligence techniques at different stages. The evaluation results using public datasets showed that Deep Learning based detections were highly successful compared with baseline models.  
\end{abstract}

\section{Introduction}
\label{intro}
The majority of cybercrime has always been financially motivated, and banking Trojans or botnets have been some of the primary drivers of botnet traffic and malicious activities[1][2][5][20].  For example, The GameOver ZeuS (GOZ) group, was a crime ring that focused on corporate banking account takeovers, with an estimated 100 million dollars of losses attributed to the group. Variants of GameOver Zeus evolved even after the technical and legal takedown of the infrastructure of GOZ peer-to-peer network in 2014. 

Since the top banking botnets and takedown efforts in 2014 and 2015, researchers [21] [22 ]have observed cybercriminals learning from past experience and quickly adapting to more sophisticated technologies commonly seen in advanced persistent threat (APT) attacks. Instead of stealing credentials by infecting banking customer’s computers, cybercriminals as reported in [22] directly targeted organizations, banking networks using APT techniques for financial gain as opposed to espionage.

Financial organizations have been utilizing Cyber Kill Chain (CKC) Taxonomies to support defensive and investigative tactics for analysts and experts in organizations to perform their day-to-day tasks against APT. CKC is based on the kill chain tactic of the US military’s F2T2EA (Find, Fix, Track, Target, Engage and Assess) [4] and Cyber Kill Chain (CKC) is one of the most widely used operational threat intelligence models to explain intrusion campaign activities in seven stages. Authors in [5] further proposed a CKC-based taxonomy for Banking Trojans in supporting security experts on the banking/financial industry sector. Basing on validation by using the 127 Trojan samples collected from a real-world banking environment in the UK [5].

The steps below describe in greater detail how APT-based banking trojan typically works:
\begin{enumerate}
    \item {\bf Reconnaissance and Weaponization}: Gathering information and preparation of an attack.  Using Carbanak APT [22] as an example, Cybercriminals registered new spoofing domains to impersonate a legitimate software or tech company in later spear phishing emails claiming required software update.
    \item {\bf Delivery}: Common methods of malicious payload delivery by banking Trojans are email attachments, social engineering and drive by download through spear phishing campaigns targeting employees within the victim organizations. 
    \item {\bf Exploitation and Installation}: If an employee within the targeted organizations opened the attachments or visited the malicious websites in above spear phishing emails, the vulnerability is successfully exploited, and backdoor is installed on the victim’s system.
    \item {\bf Command and Control}: Command-and-control (C2) infrastructure plays an essential role in coordinating botnets and malware. Attackers set up C2 servers to distribute commands or harvest sensitive data from victims’ computers, or gain access to the critical systems in the victim’s infrastructure. Many sophisticated malware families contain domain generation algorithms (DGAs) to generate pseudo-random domains in bulk to evade public blacklists.
    \item {\bf Action on Objectives}: Once the attackers successfully compromised the victim's networks, especially the critical systems such as money processing services or financial accounts, attackers can now perform fraudulent transactions or cash out.
    
\end{enumerate}

In this paper, we will discuss recent research efforts that applied cutting edge technologies including Big Data, Deep learning and Graph analysis. Specifically, we derive our core design goals from the following considerations: 
\begin{itemize}

    \item Based on the latest Mandiant report (M-Trends 2018) [6], the global median dwell time from compromise to discovery is up from 99 days in 2016 to 101 days in 2017, and up to a median of 498 days for some regions.  
    \item A Cyber Data Lake automatically collects, normalizes and archives hundreds of commonly used security datasets, intelligent threat inputs, etc., on top of big data/analytical ecosystems, which makes it possible to leverage different datasets and detect botnets at different stages.
    \item Advanced Artificial Intelligence (AI) techniques, such as Deep learning, Graph analysis, play a more significant role in reducing the time and cost of manual feature engineering and discovering unknown patterns for Cyber security analysts.
\end{itemize}

We present here an AI-based, early warning, multi-stage system that detects malicious Trojan activities from internal and external sources, through the whole lifecycle of the banking botnets, even ahead of the actual spear phishing campaign. Our proposed system leverages Cyber Data Lakes as a central platform housing multiple data sources, as well as an analytics platform for the detection engines; The evaluation results using public datasets showed that Deep Learning based detections were highly useful in terms of challenges such as false positive reduction, etc.

The rest of the paper is organized as below. An overview of system design is discussed in Section \ref{system_design}. System details and evaluation results are given in Sections \ref{system_details}  and \ref{eval_result}. Finally Section \ref{Conclusion} concludes the paper with our contributions and future work.

\section{System design overview}
\label{system_design}
The system design overview is based on timelines of typical banking botnets activities  corresponding to detection stages and their associated attack vectors. Moreover, the proposed detection framework is designed for any Cyber Data Lake.

\subsection{Attack Timeline and Proposed Staged Detection}

As discussed in Section \ref{intro}, a typical APT-based banking Trojan requires multiple steps to achieve its goal and the global median dwell time from compromise to discovery is 101 days, or even longer. 

We summarized the typical attack timeline together with proposed corresponding detection stages in Figure 1:
\begin{itemize}
    \item {\bf Early-Warning Detection}: Fuzzy domain spoof detection for signs of spear phishing campaigns against financial organizations as rapidly as new domains are observed.
    \item {\bf Spear Phishing Detection}:  Deep Learning based detection for spear phishing email campaigns as one way of infection.
    \item {\bf DGA Detection}:  Deep Learning based detection of indicators of infected hosts reaching out to C2 servers.
    \item {\bf Advanced data exfiltration detection}: Anomaly detection for potential data exfiltration.
\end{itemize}

\begin{figure}[hbt!]
  \centering
  \includegraphics[width=\linewidth]{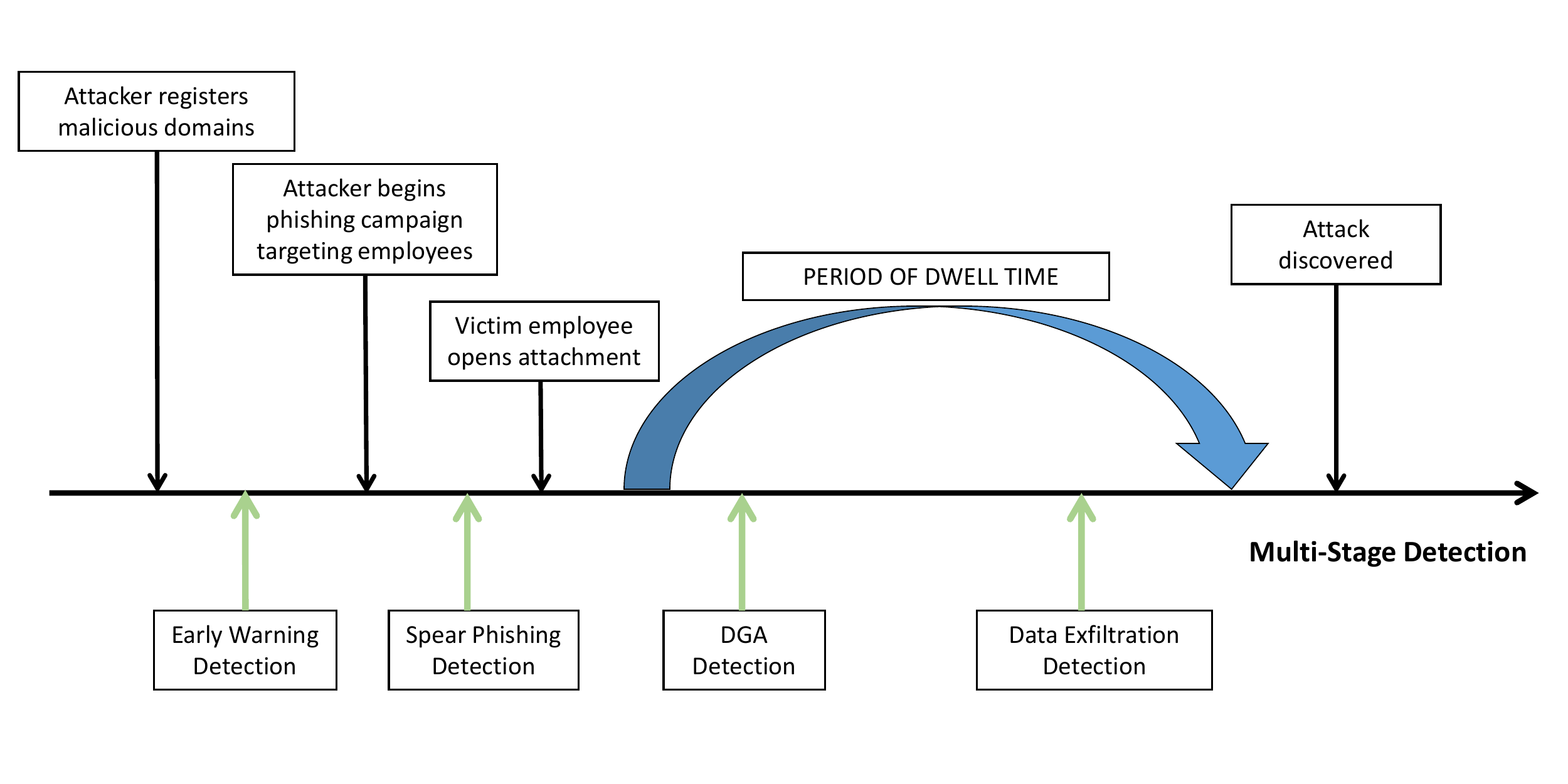}
  \caption{Attack Timeline vs. Detection Stages}
  \label{fig_timeline}
\end{figure}

\subsection{Attack Vectors}
Cyber Threat Hunting is an effective method for identifying “Delivery Channels” for “Attack Vectors” that have evaded traditional security defenses. For more complete knowledge base of adversary tactics and techniques, please refer to MITRE ATT\&CK [23]. 

Main attack vectors and delivery channels used in the multi-stage detection in Figure \ref{fig_timeline}, are outlined below:
\begin{itemize}
    \item {\bf Known Fuzzy Logics of Domain Spoof Permutations}: When attackers are registering spoof domains used in spear phishing campaigns, they use a wide range of well-known domain fuzzing permutations [7]; e.g.  {\it Homoglyph} which replaces a letter in the domain name with letters that look similar (such as amaz0n.com) and {\it Ommission} which removes one of the letters from the domain name (such as amaon.com).
    \item {\bf WHOIS Features}: When analyzing potentially malicious domains, another useful attack vector for Cyber criminals, is to bulk register domains within a short interval of time. Domains registered together may be similar to one another so that performing WHOIS queries and link analysis of detected domains are effective approaches for security analysts to spot potential spear phishing campaigns.
    \item {\bf Lexical Features}: Spear Phishing URLs and domains generated using Domain Generation Algorithms (DGAs) are lexically different from legitimate sites; which makes it possible to classify phishing URLs and DGAs by URL or domain, without any other contextual information.
    \item {\bf Traffic Patterns}: Using data exfiltration as an example, one novel way to defeat traditional network security countermeasures is called DNS tunneling.  Exploiting the simplification of DNS, an attacker can evade detection by employing tunneling technology. However, the attack traffic usually differs from normal traffic by the volume of DNS requests within a certain time frame, the payload of the DNS requests, etc. Traffic patterns can be exploited by security professionals to identify ongoing DNS tunnels. 
\end{itemize}

\subsection{Detection Framework on the Cyber Data Lake}
The Cyber Data Lake is typically used as the central platform housing multiple data components, as well as an analytics platform for multistage detection engines and serving data consumers through different methods, as shown in Figure 2.
\begin{figure}[hbt!]
  \centering
  \includegraphics[width=\linewidth]{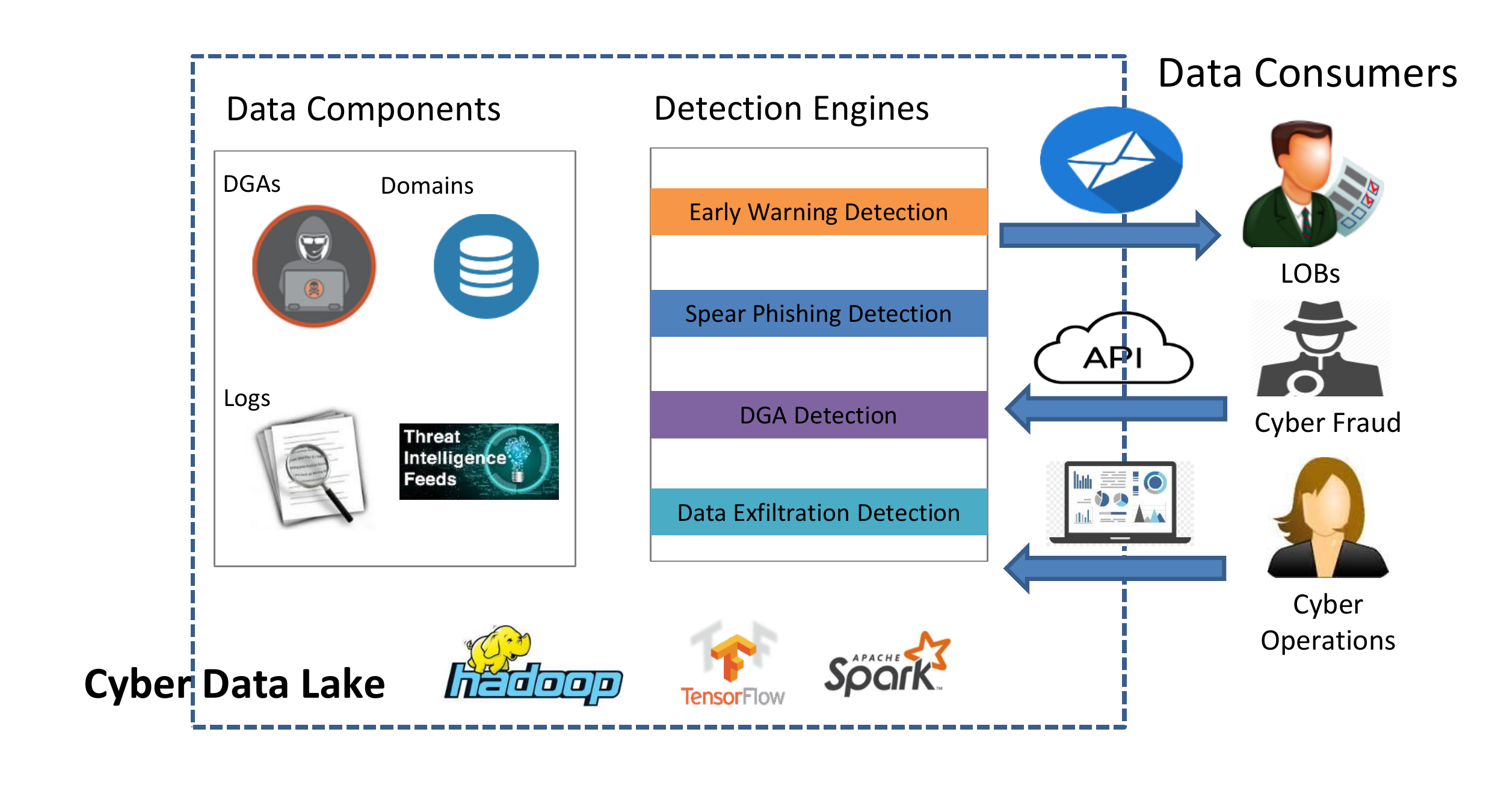}
  \caption{Multi-stage detection framework on cyber Data Lake}
  \label{fig_datalake}
\end{figure}

Data Components covered in this paper:
\begin{itemize}
    \item Realtime feed of new domains registrations and activities.
    \item Known benign and malicious domains, URLs for training/testing of the lexical model.
    \item Known rules of domain spoof permutations.
    \item Internal logs, Threat intelligence.
\end{itemize}

Analytic technologies covered in this paper:
\begin{itemize}
    \item Big data ecosystem and tools (Hadoop, Spark, etc.)
    \item Deep learning platform (Tensorflow/Keras, etc.)
    \item WHOIS/registrar information, such as registrant/email/contact, etc.
    \item Python libraries.
\end{itemize}

Furthermore,the Cyber Data Lake also provides dashboards, notebook visualizations and further investigation.

\section{System Details}
\label{system_details}

This section provides details on how each proposed detection module works, based on the corresponding input datasets and detection techniques for the specific stage of APT-based banking Trojans.

\subsection{Early Warning}

For certain APT-based banking Trojans, researchers found the malicious actors gain entry to an employee’s computer by utilizing spear phishing techniques to install a backdoor, granting them remote access to the system in order to exfiltrate data. Researchers also observed that APT domains used in the spear phishing emails were constructed in a similar lexical fashion. One of the spoofing techniques often leveraged is the impersonation of a legitimate software or tech company in an email claiming a required software update.  In the multi-stage detection system, we proposed Early Warning detection which designed to detect the spoof domains as early as they are observed (Figure 3).  The detection techniques are based on known fuzzy logics of domain spoof permutations (such as {\it Homoglyph}, {\it bitsquatting}) provided in [7]. Further investigations of detected domains could be provided by performing WHOIS queries, link analysis (e.g. HITS [8] or PageRank [9]) between domains and WHOIS info, and graph visualization. This makes it easier for security analysts to spot potential spear phishing campaigns behind the scene and reveal relationships/patterns which are unknown through other traditional analysis.  

\begin{figure}[h!]
  \centering
  \includegraphics[width=\linewidth]{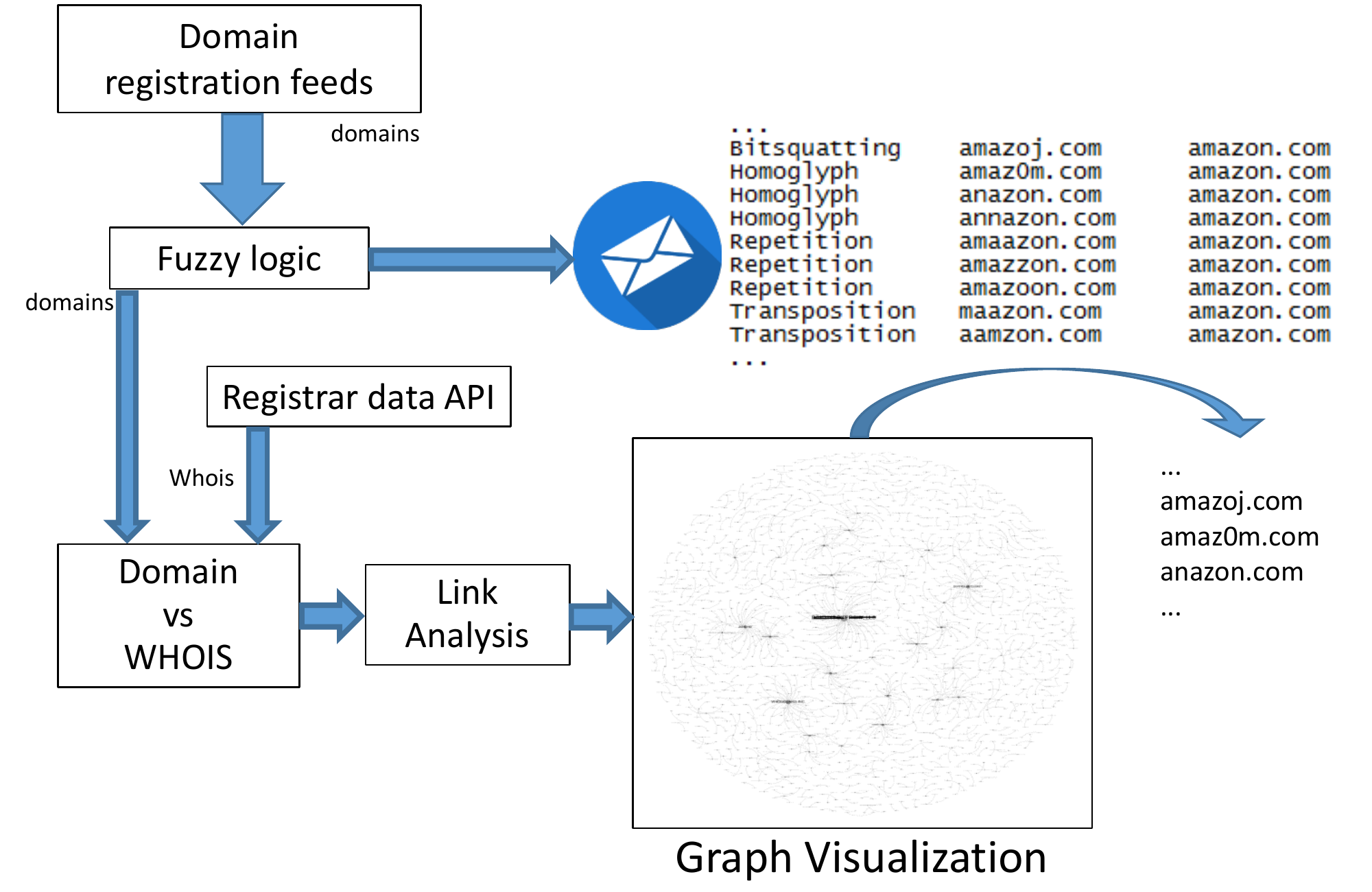}
  \caption{Early warning based on newly observed domains }
  \label{fig_ew}
\end{figure}

\subsection{Deep Learning for DGA Classification and Spear Phishing URL Classification}
Employees' computers may have been infected during a visit to a compromised website through Spear Phishing campaigns. These Spear Phishing sites are lexically different from benign sites:
\begin{itemize}
\item Mismatch URL if we read the URL from the right hand side to the left 
\item Poor spelling and grammar 
\item Asking for personal credentials (account, login, etc.)
\item Pretending as legitimate domains
\item With legitimate URLs appended at the end
\end{itemize}

In [10] [11], the authors make predictions from lexical and host-based features of URLs without examining the actual content of Web pages. 

Many sophisticated malware families utilize Domain Generation Algorithms (DGAs) to generate pseudo-random domains in bulk. The malware then attempts to connect to all or a portion of these generated domains in hopes of finding a command and control (C2) server from which it can receive instructions to pursue malicious activities. Authors in [12] presented a feature-less, per domain basis classifier based on Long Short-Term Memory networks (LSTMs), using only the domains’ names to classify DGAs.

In this paper, we applied Long Short-Term Memory networks (LSTMs) to classify both Spear Phishing URLs and DGAs under below considerations: 
\begin{enumerate}
\item Spear Phishing URLs and DGAs are lexically different from the legitimate ones where we collected a large amount of malicious as well as benign examples  through Threat intel on the Cyber Data Lake; 
\item We used LSTM units to build a model that views a URL/Domain as a character sequence and predicts whether the URL/Domain corresponds to a case of phishing/DGA;
\end{enumerate}

Following [12], the LSTM architecture is illustrated in Figure 4. 

\begin{figure}[h!]
  \centering
  \includegraphics[width=\linewidth]{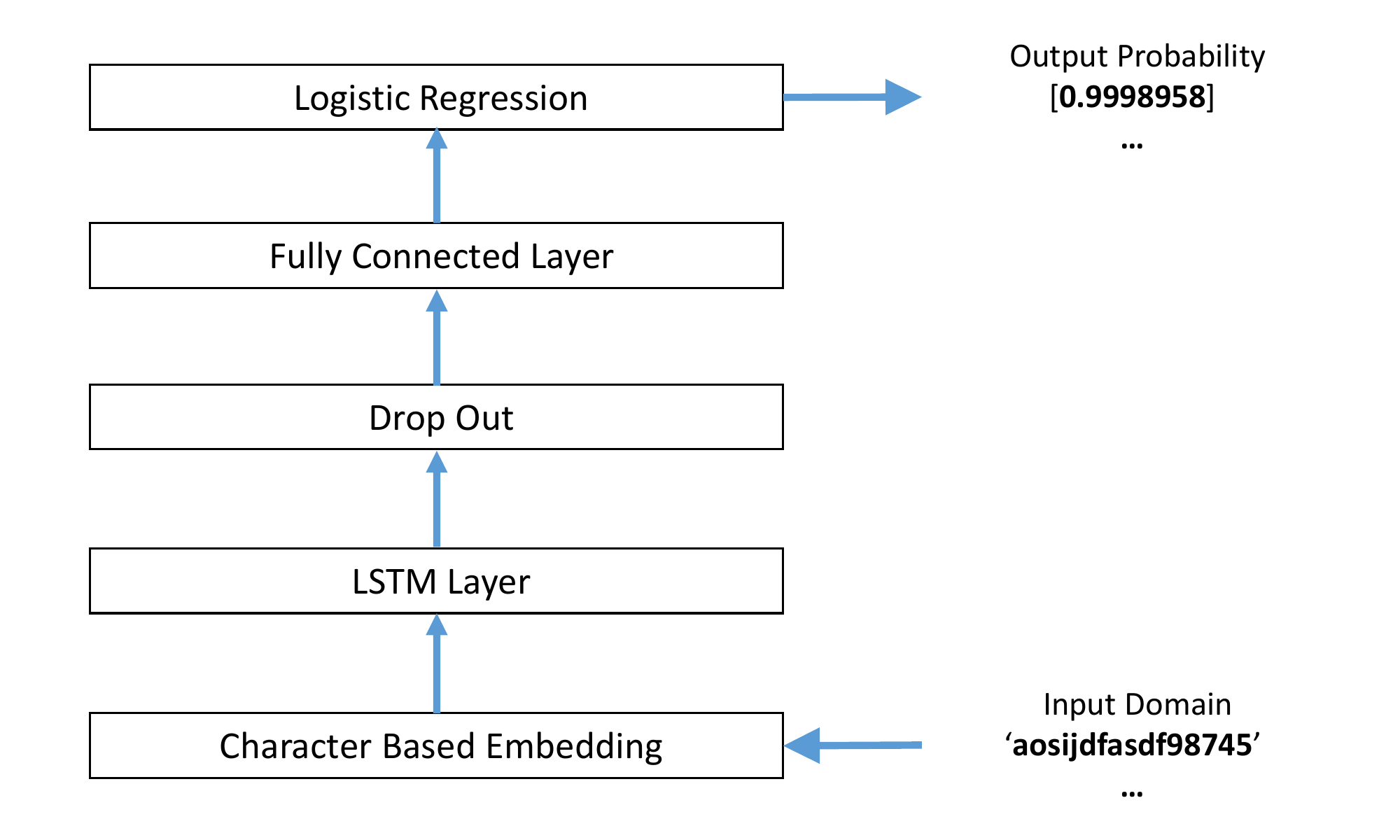}
  \caption{LSTM architecture }
  \label{fig_lstm}
\end{figure}

Each input character is embedded within a 128-dimensional vector space. The translated embedding vector is then fed to an LSTM layer as a 128-step sequence. Finally, classification is performed using a sigmoidal transfer function to an output neuron. The network is trained by backpropagation using a cross-entropy loss function and a dropout layer is added to prevent over fitting.

\subsection{DNS Tunneling Anomaly Detection}

As previously mentioned, DNS Tunneling traffic differs from usual traffic in a number of ways.  Based on these differences, we have developed a DNS tunneling detection tool to identify hidden DNS tunneling attack. To do so, we extracted a number of traffic features, included
\begin{enumerate}

    \item DNS packet length
    \item DNS request frequency
    \item DNS payload analysis:
    \begin{itemize}
     
        \item Host name length
        \item Number of unique characters
        \item Length of maximum consecutive consonants and numeric
        \item Entropy
        \item Source/destination ports
        \item Query type
    \end{itemize}

\end{enumerate}

The above features are assigned different weights similar to [13][14], and the final scores are used to identify potential attacks. The detected potential attacks are then fed to the network operation team for further investigation. Whitelist and blacklist are also applied in operation system to further reduce the noises.

\section{Evaluation Results}
\label{eval_result}
In this section, we are going to evaluate two Deep Learning models, namely, DGA classification and Phishing URL classification, using public benign datasets and public malicious threat Intel datasets. In our original work, we also used internal data for training and evaluation. However, for the purposes of this paper, only public datasets were used for the experiments.

The experiments used below four public data sources:
\begin{enumerate}

    \item Alex Top 1 million known-benign domains [15] are used as benign domains for DGA classifier evaluation.
    \item 30 DGA families [16] used to generate $\sim$ 1 million malicious domains for DGA classifier evaluation.
    \item Nonphishing URLs ( 300,000 URLs) are drawn from sources such as the DMOZ Open Directory Project [17] for Phishing URL classifier evaluation.
    \item Phishing URLs ( 267,418 URLs) are drawn from sources such as PhishTank[19] for Phishing URL classifier evaluation.

\end{enumerate}

All evaluation experiments were ran on one of the nodes from an internal GPU farm, with 4 16GB GPUs installed. Experiment source codes were written in Tensorflow/Keras, with hardware LSTM optimization which delivered up to 6x speedup. 

For DGA classification, we evaluated an LSTM model against a baseline model. The baseline model selected was the ngram (n=2) model and logistic regression classifier. The LSTM model is defined in Figure 4, with Dropout (0.5). Benign domains are coming from Alexa top 1 million domains and the DGAs are generated by 30 DGA families.
  
For Phishing URL classification, the baseline model was a stacked “bag of words” and logistic regression classifier. The LSTM model is the same as defined in Figure 4. However the maximum input length is increased to 128 for URLs comparing with 64 for domains.  Benign URLs were retrieved from DMOZ Open Directory Project and whilst phishing URLs were drawn from OpenPhish and PhishTank.

All labeled datasets were further split by training and testing datasets. For testing datasets, ROC (Receiver Operating Characteristic) curves, AUC (Area Under the curve) as well as testing accuracy were used for evaluating binary classification. Early-stopping was applied to prevent overfitting and we observed training history, e.g. Figure 5, during the hyperparameter tuning and training phases, to ensure that LSTM was not overfitting. 

\begin{figure}[h!]
  \centering
  \begin{subfigure}[b]{0.4\linewidth}
    \includegraphics[width=\linewidth]{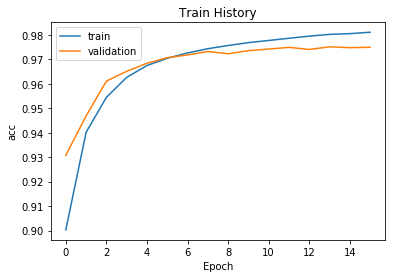}
    \caption{Accuracy on Train and Validation}
  \end{subfigure}
  \begin{subfigure}[b]{0.4\linewidth}
    \includegraphics[width=\linewidth]{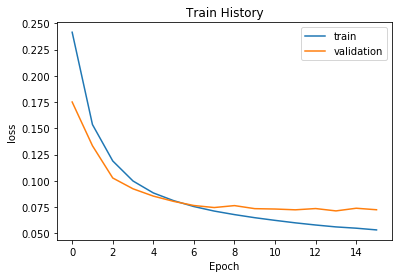}
    \caption{Loss on Train and Validation}
  \end{subfigure}
  \caption{Training history }
  \label{fig:train_history}
\end{figure}

\begin{table}
  \caption{Comparison of AUC (Area Under the curve) among 4 models}
  \label{table1}
  \centering
  \begin{tabular}{lll}
    \toprule
    Model Name     & AUC\\
    \midrule
    DGA-baseline         &0.9610       \\
    DGA-LSTM     &0.9965       \\
    Phishing-baseline     &0.9687         \\
    Phishing-LSTM     &0.9956          \\
    \bottomrule
  \end{tabular}
\end{table}

\begin{table}
  \caption{True Positive Rates vs. False Positive Rates}
  \label{table2}
  \centering
  \begin{tabular}{llllll}
    \toprule    
    Model Name     & True Positive Rate & False Positive Rate & True Positive Rate & False Positive Rate\\
    \midrule 
    DGA-baseline  &0.97 	&0.35	&0.91	&0.08    \\
    DGA-LSTM   &0.98 	&0.016	&0.91	&0.007    \\        
    \bottomrule
  \end{tabular}
\end{table}

Table 1 summarizes comparison between LSTM models with baseline models for both DGA and Phishing URL classification. LSTM model achieved higher AUC (0.9965) compared to AUC (0.9610) for the corresponding ngram model respectively.  Similarly, for Phishing URL detection, LSTM attained AUC of 0.9956, compared with 0.9687 provided by the baseline model. Both LSTM models achieved about 3\% improvement comparing with baseline models. 

More importantly, for the same given detection rate, LSTM model achieved 10+ times false positive reduction. As an example, based on Table 2, we can see that the LSTM model can classify 91\% DGA domains with 0.7\% of false positive rate while the ngram model is having 8\% of false positive rate with the same detection rate. And in Cybersecurity, false positive reduction is very critical due to the large volume of traffic on daily basis.

\section{Conclusion}
\label{Conclusion}
In this paper, we demonstrated a systematic design and implementation of banking botnet detection. Our contributions are three folds. Firstly, we proposed to use Cyber Data Lake as a common platform for both data collection and botnet detection.  Secondly, with the defense in depth mindset, we detect botnets over multiple stages of attack, ranging from early warning detection, DGA detection, Spear Phishing detection, and DNS Tunneling. Thirdly, cutting edge AI technologies such as LSTM and Graph Analysis provide a critical edge in detecting new attacks over more traditional security systems.

For future work, we will experiment with other Deep Learning techniques, such as Convolutional Neural Networks (CNN), Gated Recurrent Units (GRU), or their combinations to further enhance DGA and Phishing URL detection. And we will continue research on further operational challenges, such as noise reduction in real networks.


\section*{References}

\medskip

\small

[1] Tiirmaa-Klaar, H., \ Gassen, J., \ Gerhards-Padilla, E., \& Martini, P. \ (2013) Botnets: How to Fight the Ever-Growing Threat on a Technical Level. In {\it Botnets}. pp. 41–97. London: Springer. 

[2] Subrahmanian, V.S., \ Ovelgonne, M., \ Dumitras, T. \& Prakash, B.A. \ (2015) {\it The Global Cyber-Vulnerability Report}, November Ed. Springer International Publishing.

[3] Goznym Banking Trojan \ (2016). {\it McAfee Labs Threat Advisory}, May 23 Ed. Online: \url{https://kc.mcafee.com/resources/sites/MCAFEE/content/live/PRODUCT_DOCUMENTATION/26000/PD26519/en_US/McAfee Labs Threat Advisory - Goznym_Banking_Trojan.pdf}.

[4] Hutchins, E.M., \ Clopp, M.J. \& Amin, R.M. \ (2014) Intelligence-driven Computer Network Defense Informed by Analysis of Adversary Campaigns and Intrusion Kill Chains. White paper. Lockheed Martin Corporation, July Ed, pp. 1–14.
 
[5] Kiwia, D., \ Dehghantanha, A., \ Choo, K.R. \& Slaughter, J. \ (2017) A Cyber Kill Chain Based Taxonomy of Banking Trojans for Evolutionary Computational Intelligence, {\it Journal of Computational Science}, 27, pp. 394-409. Online: \url{https://doi.org/10.1016/j.jocs.2017.10.020}

[6] {\it Fireeye Mandiant M-Trends 2018 report}. Online: \url{ https://www.fireeye.com/content/dam/collateral/en/mtrends-2018.pdf}.

[7] Dnstwist. Online: \url{https://github.com/elceef/dnstwist}

[8] Ding, C., \ He, X., \ Husbands, P., \ Zha, H. \& Simon, H. \ (2004) Link Analysis: Hubs and Authorities on the World. {\it Siam Review}, 46(2), pp. 256-268.

[9] Page, L., \ Brin, S. \ Motwani, R. \& Winograd, T. \ (1999) The Pagerank Citation Ranking: Bringing order to the Web. Technical report, Stanford Digital Libraries SIDL-WP-1999-10120, 1999.

[10] Ma , J., \ Saul, L.K., \ Savage, S. \& Voelker, G.M. \ (2011) Learning to detect malicious URLs, {\it ACM Transactions on Intelligent Systems and Technology (TIST)}, 2(3), pp.1-24.

[11] Ma, J., \ Saul, L.K., \ Savage, S. \& Voelker, G.M. \ (2009) Beyond Blacklists: Learning to Detect Malicious Web Sites from Suspicious URLs, {\it World Wide Web Internet And Web Information Systems}, pp. 1245-1253. 

[12] Woodbridge, J., \ Anderson, H.S., \ Ahuja, A. \& Grant, D. \ (2016) Predicting Domain Generation Algorithms with Long Short-Term Memory Networks. Online: \url{ http://arxiv.org/abs/1611.00791}. 

[13] Farnham, G. \& Atlasis, A. \ (2013) Detecting DNS Tunneling. Online: \url { https://www.sans.org/reading-room/whitepapers/dns/paper/34152}.

[14] Jaworski, S. \ (2018) Using Splunk to detect DNS Tunneling. Online: \url{https://www.sans.org/reading-room/whitepapers/malicious/paper/37022}.

[15] Does Alexa have a list of its top-ranked websites? Online: \url{https://support.alexa.com/hc/en-us/articles/200449834-Does-Alexa-have-a-list-of-its-top-ranked-websites-}.

[16] Domain Generation Algorithms. Online: \url{https://github.com/baderj/domain_generation_algorithms}.

[17] DMOZ Open Directory Project (2016) Online: \url{ http://www.dmoz.org}.

[18] Whois Lookup Online: \url{https://whois.net/}.

[19] PhishTank (2018) Online:\url{http://www.phishtank.com}.

[20] Top 10 Banking Trojans for 2017: What You Need to Know (2017) Online:\url{https://blog.barkly.com/top-banking-trojans-2017}.

[21] Banking Botnets: The Battle Continues, Dell SecureWorks Counter Threat Unit™ Threat Intelligence, February 19, 2016 Online:\url{https://www.secureworks.com/research/banking-botnets-the-battle-continues}.

[22] The Great Bank Robbery: the Carbanak APT, Version 2.0, Kasperskey, February, 2015, Online:\url{https://krebsonsecurity.com/wp-content/uploads/2015/02/Carbanak_APT_eng.pdf}.

[23] MITRE ATT\&CK, Online:\url{https://attack.mitre.org/}.

\end{document}